\documentclass[
superscriptaddress,
nofootinbib,
twocolumn,
 amsmath,amssymb,
 aps,
pra,
notitlepage,
longbibliography
]{revtex4-1}

\usepackage{dcolumn}% Align table columns on decimal point
\usepackage{bm}% bold math
\usepackage{epsfig}
\usepackage{graphicx}
\usepackage{latexsym}
\usepackage{amsfonts}
\usepackage{setspace}
\usepackage{graphicx}
\usepackage{amsmath}
\usepackage{verbatim}
\usepackage{color}
\usepackage{SIunits}
\usepackage{hyperref}
\usepackage{braket}
\usepackage{mathrsfs}
\newcommand{\be}{\begin{equation}}
\newcommand{\ee}{\end{equation}}
\newcommand{\ba}{\begin{array}}
\newcommand{\ea}{\end{array}}
\newcommand{\bqa}{\begin{eqnarray}}
\newcommand{\eqa}{\end{eqnarray}}
\newcommand{\bd}{\bold}

\begin{document}

\title{Anomalous quantum Hall effect of light in Bloch-wave modulated photonic crystals}

\author{Kejie Fang} 
\email{kfang3@illinois.edu}
%\homepage{https://fang.ece.illinois.edu}
%\thanks{These authors contributed equally to this work.}
\affiliation{Department of Electrical and Computer Engineering, University of Illinois at Urbana-Champaign, Urbana, IL 61801 USA}
\affiliation{Micro and Nanotechnology Laboratory, University of Illinois at Urbana-Champaign, Urbana, IL 61801 USA}
\author{Yunkai Wang} 
\affiliation{Micro and Nanotechnology Laboratory, University of Illinois at Urbana-Champaign, Urbana, IL 61801 USA}
\affiliation{Department of Physics, University of Illinois at Urbana-Champaign, Urbana, IL 61801 USA}

\begin{abstract} 

Effective magnetic fields have enabled unprecedented manipulation of neutral particles including photons. In most studied cases, the effective gauge fields are defined through the phase of mode coupling between spatially discrete elements, such as optical resonators and waveguides in the case for photons. Here, in the paradigm of Bloch-wave modulated photonic crystals, we show creation of effective magnetic fields for photons in conventional dielectric continua for the first time, via Floquet band engineering. By controlling the phase and wavevector of Bloch waves, we demonstrated anomalous quantum Hall effect for light with distinct topological band features due to delocalized wave interference. Based on a cavity-free architecture, in which Bloch-wave modulations can be enhanced using guided-resonances in photonic crystals, the study here opens the door to the realization of effective magnetic fields at large scales for optical beam steering and topological light-matter phases with broken time-reversal symmetry.

\end{abstract}
%\pacs{}

\maketitle

Photons, being charge neutral, are not susceptible to magnetic fields. Recently, several methods have been proposed to create effective magnetic fields for photons, including chiral mode coupling  \cite{hafezi2011robust} and dynamic index modulation \cite{fang2012realizing}, leading to topological photonic states \cite{hafezi2013imaging,lin2016photonic} and nonreciprocal light propagation \cite{tzuang2014non,fang2013controlling}. In the scheme of dynamic index modulation, the effective gauge field is equivalent to the phase of a point modulation that is exerted to mediate the coupling between two spatially localized optical resonators with different frequencies \cite{fang2012realizing}. Such a revelation enables analogies between modulated optical resonator lattices and condensed matter systems under magnetic fields via commonly used tight-binding models, which also imposes challenges for experimental realization and practical uses. However, it is unknown how to create effective magnetic fields for photons in a continuum of conventional dielectrics, where electromagnetic fields are delocalized, invalidating the notion of local phase of mode coupling for effective gauge fields. Here, we study a new paradigm of dynamically modulated continua, that is photonic crystals subject to Bloch-wave modulations, in which spatial gauge fields for photons can be created via Floquet band engineering \cite{kitagawa2010topological,lindner2011floquet}. In this approach, the continuum modulation induces static-band hybridization, leading to an equation of motion for electromagnetic waves that resembles that of charged particles under magnetic fields.

The paradigm of modulated electromagnetic continuum not only extends the concept of effective magnetic fields for photons to a largely unexplored yet experimentally accessible regime, but also reveals topological photonic effects that have not been demonstrated before. As we will show, by selecting the wavevectors of Bloch-wave modulations, net effective magnetic flux through the unit cell of photonic crystals vanishes. Nevertheless, the Floquet bands can still attain nonzero Chern numbers in the presence of time-reversal symmetry breaking caused by the dynamic modulation. This result represents the first anomalous quantum Hall effect for light in Floquet engineered photonic systems. 

We developed a first-principle based formalism along with \emph{ab initio} simulations to reveal unique topological band features in Bloch-wave modulated photonic crystals due to delocalized wave interference. We also propose to use guided resonances or bound states in the continuum \cite{hsu2013observation, zhao2019} to enhance the strength of Bloch-wave modulations that can be readily implemented with highly-transducing optical or acoustic pump fields \cite{fuhrmann2011dynamic, sohn2018time}. As a result, the proposed paradigm of modulated continuum here opens the door to large-scale realization of effective magnetic fields for photons in  normal dielectrics for new types of beam steering and topological states with broken time-reversal symmetry.

To study photonic crystals under continuum modulations, we start from Maxwell's equation with isotropic and temporally-periodic permittivity $\epsilon(\bd r, t)=\epsilon(\bd r)+\delta(\bd r,t)$:
\be\label{Tmaxwell}
 i\frac{\partial}{\partial t} \left( \ba{c}   \epsilon(\bd r,t)\bd E \\  \bd H  \ea \right)=\left( \ba{cc}  & i\nabla\times \\  -i\nabla\times  &  \ea \right)\left( \ba{c}   \bold E \\  \bd H  \ea \right).
\ee
Here, $\epsilon(\bd r)$ is the spatially-periodic permittivity which defines the static photonic crystal, and $\delta(\bd r,t)=\delta(\bd r)\textrm{cos}\big(\omega t+\phi(\bd r)\big)$ is the temporal modulation of the permittivity with frequency $\omega$, amplitude $\delta(\bd r)$, and phase $\phi(\bd r)$, all of which are real. We have set $\epsilon_0=\mu_0=1$ and $\mu=1$ as for most dielectric materials at optical frequencies. Because of the time-periodicity of the permittivity, according to Floquet theorem \cite{chicone2006ordinary}, the eigenmodes of Eq. \ref{Tmaxwell} can be found by decomposition of the fields into harmonics of the modulation frequency, i.e., $(\bd E \quad \bd H)^T\equiv\Psi=\sum_{n=-\infty}^\infty \psi_n e^{-i\chi t+in\omega t}$, where $\chi$ is the quasi-frequency (see Supplementary Information (SI)). We coin the resulting time-independent eigenmode equation the Floquet-Maxwell equation. 

When the modulation has the form of Bloch waves, i.e., $\delta(\bd r)e^{i\phi(\bd r)}=u(\bd r)e^{i\bd q\cdot\bd r}$ and $u(\bd r)$ a periodic function (note $\phi(\bd r)$ needs not to be $\bd q\cdot\bd r$), we call such modulated dielectric structures the Floquet photonic crystal. If the periodicities of $u(\bd r)$ and the static photonic crystal are commensurable, then by applying a gauge transformation $U_{\textrm{gauge}}: \psi_n \rightarrow e^{in\bd q\cdot \bd r}\psi_n$, the Floquet-Maxwell equation becomes spatially-periodic and thus the eigenvalue $\chi$ can be labeled by a Bloch wavevector $\bd k$, forming the Floquet bandstructure (see SI). According to this formalism, the generation of Floquet bandstructure, to the leading order, can be intuitively understood as the result of modulation induced static-band hybridization after frequency and momentum shift of the bands by $\omega$ and $\bd q$, respectively. Because of the non-vanishing momentum of the modulation, the Floquet bandstructure, with infinitely repeated branches (i.e., $\chi+n\omega$, $\forall n$, is also a solution of the Floquet-Maxwell equation), has the property $\chi(\bd k +\bd q)=\chi(\bd k)-\omega$. 

With a spatially varying modulation phase $\phi(\bd r)$, time-reversal symmetry in modulated photonic crystals is explicitly broken, as the modulated permittivity is not invariant under time reversal $t \rightarrow -t$ for arbitrary positions. We use Floquet band engineering, i.e., modulation induced static-band coupling, to derive an effective gauge field for photons. For this purpose, we consider two static bands under dynamic modulations with a frequency close to the bandgap which is larger than the bandwidth of each band. In this case, one can write down an approximate coupled-\emph{band} equation as reminiscence of the Floquet-Maxwell equation,
\begin{gather}
\label{2band1}\frac{\partial^2 \bd H_1}{\partial t^2}= -\nabla\times \frac{1}{\epsilon}\nabla\times \bd H_1+\nabla\times\frac{\delta}{\epsilon^2}e^{i\phi}\nabla\times\bd H_2,\\
\label{2band2}\frac{\partial^2 \bd H_2}{\partial t^2}= -\nabla\times \frac{1}{\epsilon}\nabla\times \bd H_2+\nabla\times\frac{\delta}{\epsilon^2}e^{-i\phi}\nabla\times\bd H_1,
\end{gather}
where $\bd H_{1,2}$ are the magnetizing fields associated with the two bands. The effective gauge field for photons can be identified by comparing the equation of motion for the electromagnetic fields of one band with the Hamiltonian of charged particles in the presence of gauge fields, i.e., $\hat H=(-i\nabla-q\bd A)^2/2m$, where the gauge field is the imaginary coefficient of the term linear in $\nabla(\cdot)$. Using this method, we find 
\be\label{TEgauge}
\bd A_\mathrm{eff}(\bd r)=\frac{1}{\omega^2}\mathrm{Im}\left\{\nabla\cdot\left(\frac{\delta}{\epsilon^2}e^{i\phi}\nabla(\nabla\frac{\delta}{\epsilon^2}e^{-i\phi})\right)\right\}
\ee
for transverse electric (TE) modes (the $\cdot$ means the scalar product between the first $\nabla$ and the last $\nabla$) and 
\be\label{TMgauge}
\bd A_\mathrm{eff}(\bd r)=\frac{2}{\omega^2}\mathrm{Im}\left\{\nabla^2(\frac{\delta}{\epsilon^2}e^{i\phi}\nabla\frac{\delta}{\epsilon^2}e^{-i\phi})\right\}
\ee
for transverse magnetic (TM) modes in two-dimensional photonic crystals, respectively (see SI). 

From Eqs. \ref{TEgauge} and \ref{TMgauge}, it is remarkable that the effective gauge field is largely determined by the modulation phase, similar to the case of discretely modulated resonator lattices \cite{fang2012realizing}; it is zero if $\phi(\bd r)=\textrm{const.}$. However, in the continuum the effective gauge field can be defined in space by the modulation phase point-to-point, while in resonator lattices it is implicitly related to the modulation phase through a line integration. One also finds that for Bloch-wave modulations with $u(\bd r)$ having the same periodicity as the static photonic crystal, $\bd A_\mathrm{eff}(\bd r)$ is periodic and net effective magnetic flux through a unit cell vanishes. As we will show next, for certain Floquet photonic crystals, even though the net magnetic flux is zero through a unit cell, the Floquet bands attain nonzero Chern numbers. This represents a photonic analogue of the anomalous quantum Hall effect \cite{haldane1988model}, which is distinct from the photonic analogues of quantum Hall and quantum spin Hall effects proposed in Ref. \cite{fang2012realizing} and Ref. \cite{hafezi2011robust}, respectively.

\begin{figure*}[!htb]
\begin{center}
\includegraphics[width=2\columnwidth]{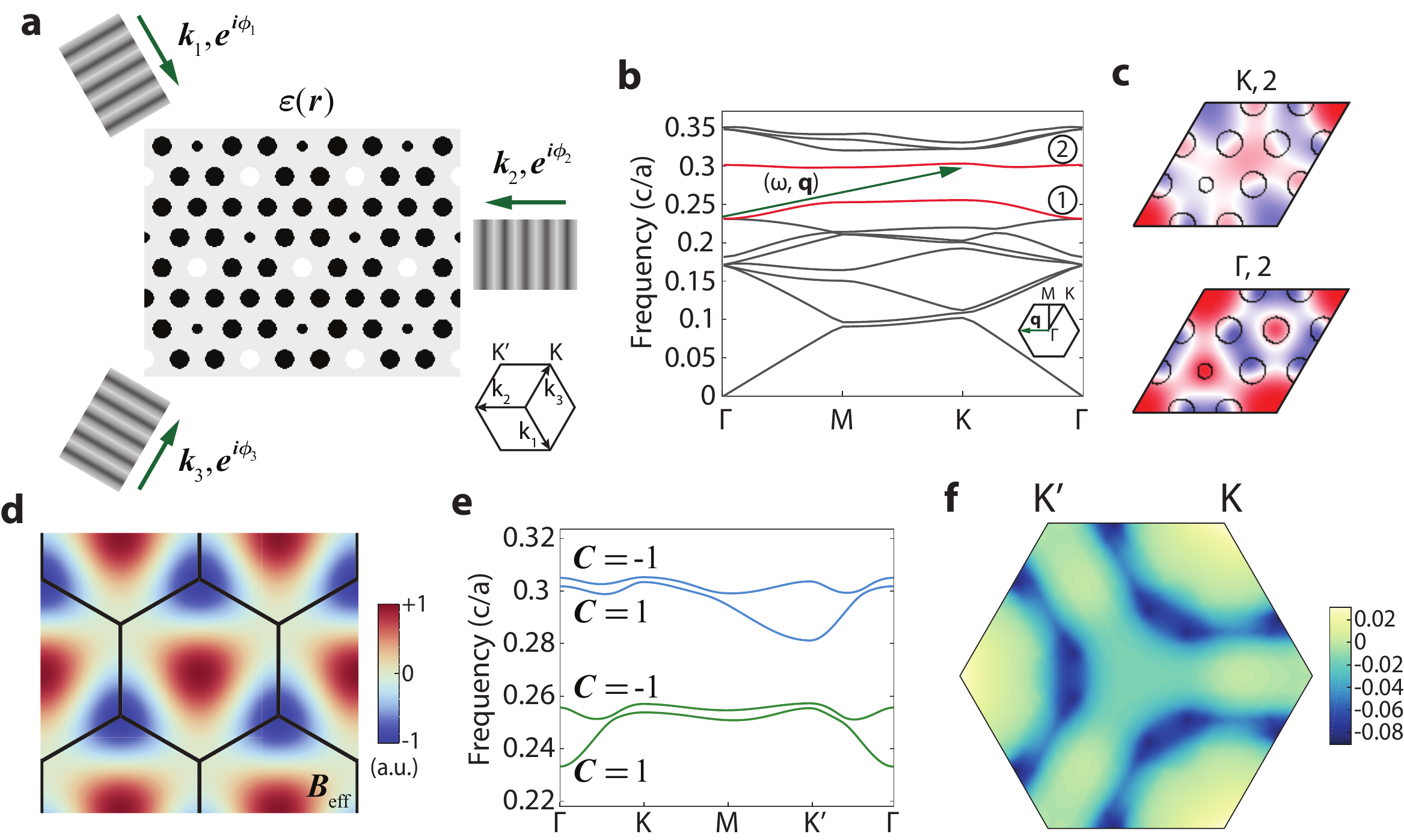}
\caption{\textbf{Effective magnetic fields in the Floquet photonic crystal.} \textbf{a}, The Floquet photonic crystal consists of a static photonic crystal and permittivity modulations of three Bloch waves with momenta equal to the three $K$ points and phases $\phi_{1,2,3}$. Black, gray, and white areas in the static photonic crystal have relative permittivity 12, 2, and 1, respectively. \textbf{b}, Bandstructure of the TM modes of the static photonic crystal. The permittivity modulation with frequency $\omega$ and Bloch momentum $\bd q$ (green arrow) is used to couple the two highlighted bands (red). \textbf{c}, Out-of-plane electric field in the unit cell corresponding to the modes of band 2 at $\Gamma$ and $K$ points. \textbf{d}, Distribution of the effective magnetic field $\bd B_{\textrm{eff}}$ induced by the Bloch-wave modulations for $\phi_1=0$, $\phi_2=2\pi/3$, and $\phi_3=4\pi/3$. The net effective magnetic flux through the Wigner-Seitz unit cell is zero. \textbf{e}, Floquet bandstructure of the modulated photonic crystal, formed from the hybridization of static bands 1 and 2 of \textbf{b}. Chern number of each band is indicated for the modulation parameters specified in the text. \textbf{f}, Berry curvature in the Brillouin zone for the top blue Floquet band in \textbf{e}. } 
\label{fig2}
\end{center}
\end{figure*}

The example of Floquet photonic crystal we studied begins with a two-dimensional photonic crystal with a triangular lattice (lattice constant $a$) of dielectric rods ($r=0.27a$) with relative permittivity $\epsilon_r=12$ (corresponding to silicon or gallium arsenide) in a background with relative permittivity $\epsilon_r=2$ (corresponding to silicon dioxide) (Fig. \ref{fig2}a). A honeycomb sub-lattice (lattice constant $3a$) is created inside the triangular lattice, consisting of smaller rods ($r=0.145a$, $\epsilon_r=12$) and holes ($r=0.27a$, $\epsilon_r=1$). The bandstructure of the TM modes calculated using plane-wave expansion method \cite{johnson2001block} is shown in Fig. \ref{fig2}b and the out-of-plane electric field of the modes of the highlighted band 2 at $\Gamma$ and $K$ points is shown in Fig. \ref{fig2}c. As time-reversal symmetry is not broken in the static photonic crystal, all the bands are topologically trivial with zero Chern numbers.  

Next we introduce temporal modulations of permittivity in order to couple bands 1 and 2 in Fig. \ref{fig2}b to create topologically nontrivial Floquet bands. The frequency of the modulation is slightly larger than the bandgap between the two bands and its spatial profile has the following form,
\be\label{mod}
\delta(\bd r)e^{i\phi(\bd r)}=\sum_{j=1}^3u_j(\bd r)e^{i(\bd k_j\cdot\bd r+\phi_j)},
\ee
where $u_{1,2,3}(\bd r)$ are functions with the same spatial periodicity as the static photonic crystal and we choose the coordinate origin at the center of an air hole. The modulation of Eq. \ref{mod} is a superposition of three Bloch waves, which could lead to rotating distribution of permittivity. For example, if we consider the case where $\bd k_{1,2,3}$ are the three equivalent $K$ points in the Brillouin zone, $u_{1,2,3}(\bd r)\equiv \bar u(\bd r)$ possessing $120^\degree$ rotational symmetry, and $\phi_j=\frac{2(j-1)\pi}{3}$, then we have $\hat R_{120^\degree}\epsilon(\bd r, t)=\epsilon(\bd r, t-2\pi/3\omega)$, where $\hat R_{120^\degree}$ is the operation of $120^\degree$ clockwise rotation in real space. As such, by adjusting the relative phase between the Bloch-waves, we can effectively circulate the permittivity and change its chirality.

Meanwhile, because $\bd k_i-\bd k_j$ are reciprocal lattice vectors, $\delta(\bd r)e^{i\phi(\bd r)}$ is by itself of the Bloch-wave form, with a Bloch wavevector $\bd q=\bd k_j$, $\forall j$. The effective magnetic field for band 2 induced by the modulation is calculated using $\bd B_{\textrm{eff}}(\bd r)=\nabla\times \bd A_{\textrm{eff}}(\bd r)$ and Eq. \ref{TMgauge} for $\bar u(\bd r)\propto\epsilon(\bd r)^2$ and $\phi_j=\frac{2(j-1)\pi}{3}$, which is shown in Fig. \ref{fig2}d. The net magnetic flux through the Wigner-Seitz cell is zero as expected. Fig. \ref{fig2}e shows the calculated TM Floquet bands (with two repeated branches) by coupling the static bands 1 and 2 with modulation parameters $\omega=0.0491\times 2\pi c/a$, $\phi_j=\frac{2(j-1)\pi}{3}$, and $\bar u(\bd r)=0.25$ (in dielectrics) or 0 (in air). We see that $K$ and $K'$ points are no longer equivalent due to the momentum-carrying modulation. Although the net magnetic flux through a unit cell is zero, we find each Floquet band has nonzero Chern number $C=\pm 1$ (see SI), with the Berry curvature of one Floquet band shown in Fig. \ref{fig2}f. This result represents a photonic analogue of the anomalous quantum Hall effect \cite{haldane1988model} realized in a structure beyond tight-binding models, and indicates the existence of one-way edge modes in the Floquet bandgap \cite{rudner2013anomalous}. 

\begin{figure}[!htb]
\begin{center}
\includegraphics[width=\columnwidth]{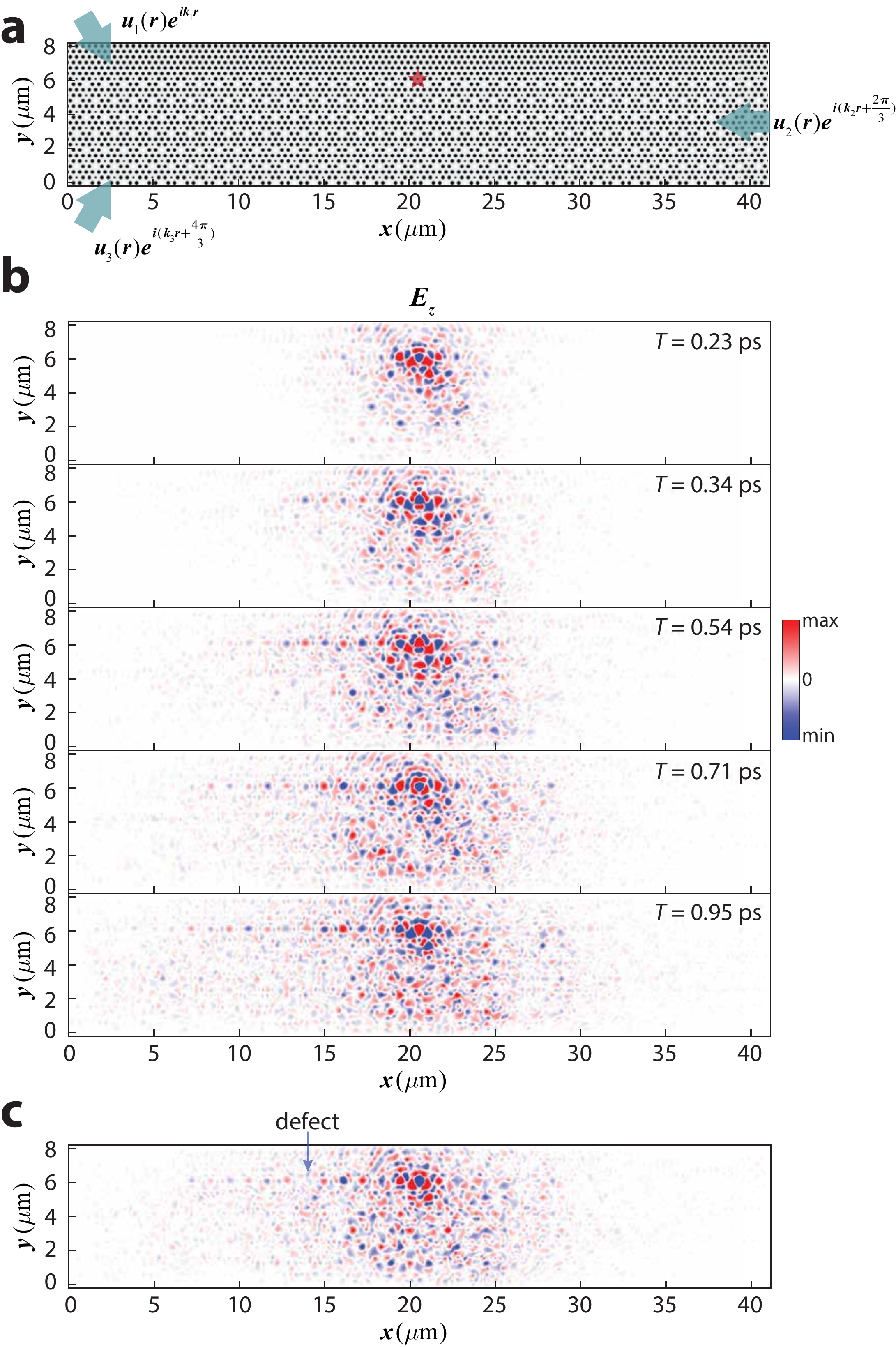}
\caption{\textbf{The anomalous topological edge mode.} \textbf{a}, The full photonic structure used in the FDTD simulation with three Bloch-wave modulations with phase lagging of $2\pi/3$. The star indicates the location of the point source.  \textbf{b}, Propagation of the electric field $E_z$ excited by the TM-polarized continuous-wave point source. A left-propagating one-way mode exists on the edge ($y\sim 6\ \mu$m) of the Floquet photonic crystal. \textbf{c}, Reflection-immunity of the one-way edge mode in the presence of a defect. Here the defect is created by removing a small dielectric rod on the edge. } 
\label{fig3}
\end{center}
\end{figure}

We numerically demonstrate the anomalous topological edge mode through finite-difference time-domain (FDTD) simulation using Maxwell's equation. The full simulation domain is shown in Fig. \ref{fig3}a with $a=0.37$ $\mu$m. The boundaries of the simulation domain are composed of perfectly-matched layers. The Floquet photonic crystal occupies part of the simulation domain with an edge at $y\sim6$ $\mu$m. The applied temporal modulation is the same as that used for the calculation of Floquet bandstructure above. A TM-polarized, continuous-wave point source with frequency $\omega_s/2\pi=0.2549c/a=206.7$ THz on the edge of the Floquet photonic crystal is used to excite the electromagnetic waves. Time-evolution of the excited electric field is shown in Fig. \ref{fig3}b, where a one-way edge mode can be clearly identified, along with bulk excitations due to the incomplete Floquet bandgap, which can be optimized by further engineering of the photonic crystal and temporal modulation. When the edge mode encounters a defect, it keeps unidirectional propagation without reflection (Fig. \ref{fig3}c), which is the evidence of topological protection. 

\begin{figure}[!htb]
\begin{center}
\includegraphics[width=\columnwidth]{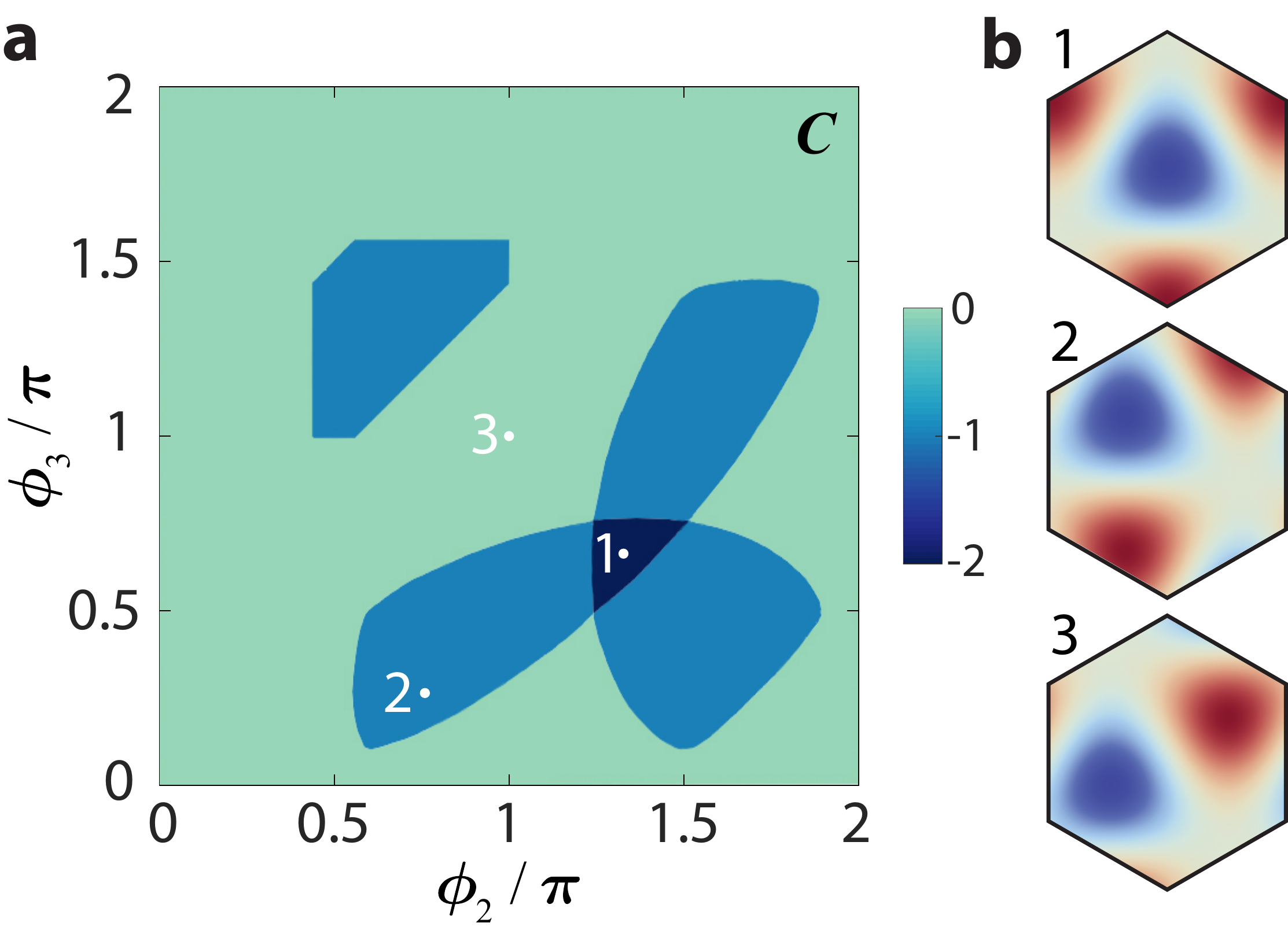}
\caption{\textbf{Phase diagram of the Floquet photonic crystal.} \textbf{a}, The Chern number of the higher frequency band of the Floquet two-band model for varying $\phi_{2}$ and $\phi_{3}$ ($\phi_1=0$). \textbf{b}, The effective magnetic field in the Wigner-Seitz cell corresponding to the phases indicated by points 1 $(4\pi/3, 2\pi/3)$, 2 $(3\pi/4, \pi/4)$, and 3 $(\pi, \pi)$. Fig. \ref{fig2}d corresponds to point $(2\pi/3, 4\pi/3)$. For the calculation of the effective magnetic field, we used $\bar u(\bd r)\propto\epsilon(\bd r)^2$ to remove the higher order texture due to the static permittivity distribution in the photonic crystal.} 
\label{fig4}
\end{center}
\end{figure}

The topological properties of Floquet photonic crystals can be controlled by the phase of Bloch waves, which also determines the distribution of the effective magnetic field in real space. Fig. \ref{fig4}a shows a phase diagram of the model above, where topologically different phases with Chern number ranging from 0 to -2 exist. The distribution of the effective magnetic field in the Wigner-Seitz cell for a few phases is shown in Fig. \ref{fig4}b. In contrary to the tight-binding model of anomalous quantum Hall effect \cite{haldane1988model}, where the sign of Chern number is related to the chirality of electron hopping amplitude, the sign of Chern number in this continuum model is not directly related to the chirality of modulations, i.e., $\phi_3>\phi_2>\phi_1$ or $\phi_2>\phi_3>\phi_1$, and the two chiralities have very different Chern number distributions. We also find that if the modulation strength is increased, positive Chern number appears around $\phi_1=\phi_2=\phi_3=0$ mod $2\pi$ (see SI), which is remarkable as intuitively the Chern number is expected to be zero when the chirality of modulation is absent. These novel topological band features in Floquet photonic crystals are largely due to the delocalized wave interference in the continuum and difficult to produce in tight-binding type of systems. 

Using Bloch waves to generate topological photonic states whose property can be simply controlled by the phase and wavevector of Bloch waves has practical significance. In photonic crystals, the Bloch-wave type of permittivity modulations can be generated through nonlinear optical effects or optomechanical interactions. The nonlinear optical approach is suitable for modulating photonic crystals with two static bands separated by an optical frequency bandgap, and the permittivity modulation, $\delta(\bd r, t)=\chi^{(2)}E(\bd r, t)$, can be generated by a pump field $E(\bd r, t)$ in materials with second order susceptibility $\chi^{(2)}$. Nonlinear optical interactions can be significantly enhanced in photonic crystals made from strong nonlinear materials \cite{combrie2008gaas,liang2017high} and using slow light effect \cite{baba2008slow}. Here, we propose to enhance the Bloch-wave modulations using guided resonances \cite{fan2002analysis} or bound states in the continuum (BICs) \cite{hsu2013observation}, which are long-lived, delocalized resonances that can have nonzero Bloch momentum in two dimensional photonic crystal slabs. As an estimation of the available permittivity modulation with optical BICs, we assume BICs with a quality factor $Q=10^6$ in a photonic crystal slab with size $V=100\ \mu \mathrm{m}^2\times 200$ nm made from GaAs ($\chi^{(2)}=120$ pm/V). Then modulation strength $\delta(\bd r)\sim~0.1$ can be achieved using a pump field with frequency $\omega/2\pi=200$ THz and mediocre power $P\approx\frac{1}{2}\epsilon_0\epsilon_rV(\frac{\delta}{\chi^{(2)}})^2 \frac{\omega}{2\pi Q}\approx150$ mW.

On the other hand, permittivity modulations induced by acoustic waves, i.e., $\delta(\bd r, t)=\nabla\epsilon(\bd r)\cdot \bd Q(\bd r, t)$, where $\bd Q(\bd r, t)$ is the mechanical displacement, is suitable for modulating photonic crystals with two bands separated by a small bandgap in the gigahertz frequency range. Thanks to the progress of opto- and electro-mechanics in recent years, large mechanical displacement can be piezoelectrically excited on microchips \cite{balram2016coherent,sohn2018time} and significant interaction between optical fields and gigahertz mechanical modes can be obtained in mechanically-compliant photonic crystals \cite{eichenfield2009optomechanical,fang2017generalized}. To achieve $\delta(\bd r)\sim0.1$, we assume that on average $|\bd Q|\sim 1$ nm might be required. Using acoustic BICs \cite{zhao2019} with a quality factor $Q=10^6$ and frequency $f=1$ GHz, and assuming typical zero-point-fluctuation $x_{\textrm{zpf}}=1$ fm in a unit cell for the gigahertz mechanical mode \cite{eichenfield2009optomechanical}, then in a photonic crystal with $N=10^4$ unit cells, the required acoustic pump power to achieve $|\bd Q|\sim 1$ nm is approximately given by $P\approx N(\frac{|\bd Q|}{x_{\textrm{zpf}}})^2\frac{hf^2}{Q}\approx 7\ \mu$W, where $h$ is the Planck constant. Here, by trapping the optical or acoustic pump fields in delocalized high-$Q$ resonances of photonic crystals, we can significantly enhance the modulation strength in a large area with experimentally available pump power, in contrast to the scheme of microwave-driven electro-optic modulations \cite{lira2012electrically,fang2012realizing}, which lacks such a mechanism because of the wavelength mismatch between the microwave and optical fields.

In summary, we have revealed the generation of effective magnetic fields for photons in photonic crystals under continuous spatio-temporal modulations. In this paradigm, we showed a photonic analogue of anomalous quantum Hall effect with unique topological band features due to delocalized wave interference. Other than that, a variety of combinations of spatio-temporal modulations and static photonic crystals can be explored. For instance, the scenario when the modulations have different periodicity than that of the static photonic crystals, e.g., when $\bd k_i-\bd k_j$ in Eq. \ref{mod} are not the primitive reciprocal lattice vectors, might be used to generate fractal photonic spectra, such as the Hofstadter butterfly \cite{hofstadter1976energy}. Higher order topological phases other than those characterized by the Chern number \cite{rodriguez2018higher} might also be realized in Floquet photonic crystals. With strong Bloch-wave modulations readily implementable on microchips, Floquet photonic crystals open the door to the realization of a plethora of optical phenomena \cite{fang2013controlling, lin2016photonic} and light-matter phases \cite{peano2016topological,ludwig2013quantum} with broken time-reversal symmetry, in a practical cavity-free architecture.

%\bibliographystyle{naturemag}
%\bibliography{./reference}

\vspace{2mm}
\noindent\textbf{Acknowledgements}\\ 
We are very grateful to Yu Shi and Shanhui Fan for providing the FDTD simulation code. This work is supported in part by US National Science Foundation under Grant No. ECCS-1809707.

\noindent\textbf{Author contributions}\\ 
K.F. proposed the theory and model. Y.W. and K.F. performed the numerical calculation. K.F. wrote the manuscript with the input from Y.W..

\onecolumngrid
\appendix

\section{Calculation of Floquet bandstructure and Chern number}

We outline the procedure for calculating the Floquet bandstructure of Bloch-wave modulated photonic crystals. Starting from Maxwell's equation with temporally-periodic and isotropic permittivity $\epsilon(\bd r, t)=\epsilon(\bd r)+\delta(\bd r,t)$:
\be\label{Tmaxwell}
 i\frac{\partial}{\partial t} \left( \ba{c}   \epsilon(\bd r,t)\bd E \\  \bd H  \ea \right)=\left( \ba{cc}  & i\nabla\times \\  -i\nabla\times  &  \ea \right)\left( \ba{c}   \bold E \\  \bd H  \ea \right),
\ee
according to Floquet theorem, we substitute scenario $(\bd E \quad \bd H)^T\equiv\Psi=\sum_{n=-\infty}^\infty \psi_n e^{-i\chi t+in\omega t}$ into Eq. \ref{Tmaxwell}, and compare the coefficients of each harmonics on the two sides of the equation, which leads to
\be\label{FloquetEigen}
 \left( \ba{ccccccc} \ddots & \ddots & \ddots &  & & &   \\   & D_{n+1}^* & C_{n+1} & D_{n+1} & & &   \\  & & D_n^* & C_n & D_n & &   \\   & & & D_{n-1}^* & C_{n-1} & D_{n-1} &    \\ & & & & \ddots & \ddots & \ddots  \ea \right)  \left( \ba{c}  \vdots \\ \psi_{n+1}\\  \psi_{n} \\ \psi_{n-1} \\ \vdots \ea \right) =\chi \left( \ba{ccccccc} \ddots & \ddots & \ddots &  & & &   \\   & B^* & A & B & & &   \\  & & B^* & A & B  & &   \\   & & & B^* & A & B &    \\ & & & & \ddots & \ddots & \ddots  \ea \right)  \left( \ba{c}  \vdots \\ \psi_{n+1}\\  \psi_{n} \\ \psi_{n-1} \\ \vdots \ea \right),
\ee
where
\begin{gather}
\nonumber A=\left( \ba{cc} \epsilon(\bd r) & 0  \\  0  & 1 \ea \right), B=\left( \ba{cc} \frac{\delta(\bd r)}{2}e^{i\phi(\bd r)} & 0 \\   0 &  0 \ea \right), \\
\nonumber C_n=\left( \ba{cc} n\omega\epsilon(\bd r) & i\nabla\times \\  -i\nabla\times  & n\omega \ea \right), D_n=n\omega B.
\end{gather}
We call Eq. \ref{FloquetEigen} the Floquet-Maxwell equation because of its derivation. 

When the modulation satisfies $\delta(\bd r)e^{i\phi(\bd r)}=u(\bd r) e^{i\bd q\cdot\bd r}$, where $u(\bd r)$ is a function with the same periodicity as the static photonic crystal, one can convert the Floquet-Maxwell equation to the following form, under the gauge transformation $\psi_n = e^{in\bd q\cdot \bd r}\tilde\psi_n$ and $M_{mn} = e^{i(m-n)\bd q\cdot \bd r} \tilde M_{mn}$ for a matrix $M$,

\be\label{gaugeFloquet}
 \left( \ba{ccccccc} \ddots & \ddots & \ddots &  & & &   \\   & \tilde D_{n+1}^* & \tilde C_{n+1} & \tilde D_{n+1} & & &   \\  & & \tilde D_n^* & \tilde C_n & \tilde D_n & &   \\   & & & \tilde D_{n-1}^* & \tilde C_{n-1} & \tilde D_{n-1} &    \\ & & & & \ddots & \ddots & \ddots  \ea \right)  \left( \ba{c}  \vdots \\ \tilde \psi_{n+1}\\  \tilde \psi_{n} \\ \tilde \psi_{n-1} \\ \vdots \ea \right) =\chi \left( \ba{ccccccc} \ddots & \ddots & \ddots &  & & &   \\   & \tilde B^* & \tilde A & \tilde B & & &   \\  & & \tilde B^* & \tilde A & \tilde B  & &   \\   & & & \tilde B^* & \tilde A & \tilde B &    \\ & & & & \ddots & \ddots & \ddots  \ea \right)  \left( \ba{c}  \vdots \\ \tilde \psi_{n+1}\\  \tilde \psi_{n} \\ \tilde \psi_{n-1} \\ \vdots \ea \right),
\ee
where 
\begin{gather}
\nonumber \tilde A=\left( \ba{cc} \epsilon(\bd r) & 0  \\  0  & 1 \ea \right), \tilde B=\left( \ba{cc} \frac{u(\bd r)}{2} & 0 \\   0 &  0 \ea \right), \\
\nonumber \tilde C_n=\left( \ba{cc} n\omega\epsilon(\bd r) & (i\nabla-n\bd q)\times \\  (-i\nabla+n\bd q)\times  & n\omega \ea \right), \tilde D_n=n\omega \tilde B.
\end{gather}
all of which are spatially periodic. As a result, the quasi-frequency $\chi$ can be labeled with a Bloch momentum $\bd k$, which forms the Floquet bandstructure. 

When the modulation is weak, we can use a perturbation method to calculate the Floquet bandstructure. It is reasonable to choose the following functions as the basis to represent the operators in Eq. \ref{gaugeFloquet},
\be\label{basis}
\psi_{m0}^\alpha(\bd k)=e^{-im\bd q\cdot \bd r}\zeta^\alpha(\bd k+m\bd q),
\ee 
where $\zeta^\alpha(\bd k+m\bd q)$ is the $\alpha-$th Bloch-phase dressed eigenmode of the static photonic crystal with Bloch momentum $\bd k+m\bd q$, which can be calculated using MIT Photonic Bands. In this basis, the operators in Eq. \ref{gaugeFloquet}, which we denote as $\mathscr{ H } \tilde\psi=\chi \mathscr { B} \tilde\psi$, is represented as $H_{mn}^{\alpha\beta}=\bra{\psi_{m0}^\alpha} \mathscr { H }_{mn} \ket{\psi_{n0}^\beta}=\int d \bd r (\bd E_{m0}^{\alpha*},\bd H_{m0}^{\alpha*})\cdot\mathscr { H }_{mn} (\bd E_{n0}^\beta,\bd H_{n0}^\beta)^T$ and $B_{mn}^{\alpha\beta}=\bra{\psi_{m0}^\alpha} \mathscr { B }_{mn} \ket{\psi_{n0}^\beta}=\int d \bd r (\bd E_{m0}^{\alpha*},\bd H_{m0}^{\alpha*})\cdot\mathscr { B }_{mn} (\bd E_{n0}^\beta,\bd H_{n0}^\beta)^T$, where the integration is calculated in the unit cell of the static photonic crystal. We then straightforwardly solve the numerical Floquet-Maxwell equation $H_{mn}^{\alpha\beta}(\bd k)c_n^\beta(\bd k)=\chi(\bd k) B_{mn}^{\alpha\beta}(\bd k)c_n^\beta(\bd k)$ to obtain the Floquet bandstructure $\chi(\bd k)$. We found $\chi(\bd k)$ is real for the models we have studied, even though the numerical matrix $H$ is not Hermitian by the usual definition, i.e., $H^{T*}\neq H$. Since the Floquet eigenstates are localized, i.e., further separated harmonics do not directly interact at least for weak modulations, we can truncate $m$ such that $m\omega$ is not much larger than the bandwidth of the bands under consideration, and ignore bands that are separated from the bands under consideration by more than $m\omega$.

The Berry curvature is calculated using $B(\bd k)=-i\left(\partial_{k_x}\left\langle \tilde\psi(\bd k) \left| \partial _ { k_y } \right| \tilde\psi(\bd k)\right\rangle-\partial_{k_y}\left\langle \tilde\psi(\bd k) \left| \partial _ { k_x } \right| \tilde\psi(\bd k)\right\rangle\right)$, where $\tilde\psi(\bd k)=(\cdots, c_m^\alpha(\bd k)\psi_{m0}^\alpha(\bd k),  \cdots)^T$ is the Floquet eigenmode of Eq. \ref{gaugeFloquet} and the inner product is calculated in the unit cell of the static photonic crystal. The Chern number is given by $C=\int d\bd k B(\bd k)/2\pi$, where the integration is calculated in the first Brillouin zone.

\section{Effective gauge field}
By comparing the wave equations of temporally modulated photonic structures to that of charged particles in magnetic fields, we can derive an effective gauge field for photons due to the temporal modulation. By making the assumption that the bandgap between two static photonic bands under modulation is larger than the bandwidth of each of them and the modulation frequency is close to the bandgap, we can write down the following coupled-mode equations from the time-dependent Maxwell's equations, using the rotating wave approximation,
\begin{gather}\label{CM1}
\frac{\partial \bd D_1}{\partial t}=\nabla\times\bd H_1,\\
\frac{\partial \bd H_1}{\partial t}=-\nabla\times\frac{1}{\epsilon}\bd D_1+\nabla\times\frac{\delta}{\epsilon^2}e^{i\phi}\bd D_2,\\
\frac{\partial \bd D_2}{\partial t}=\nabla\times\bd H_2,\\
\label{CM4}
\frac{\partial \bd H_2}{\partial t}=-\nabla\times\frac{1}{\epsilon}\bd D_2+\nabla\times\frac{\delta}{\epsilon^2}e^{-i\phi}\bd D_1,
\end{gather}
where $\{\bd D_i, \bd H_i\}$ are the fields of $i-$th band, and we have assumed $\delta\ll \epsilon$.

Eliminating the $\bd D$ fields, we obtain the coupled mode equations for the $\bd H$ fields,
\begin{gather}
\label{2band1}\frac{\partial^2 \bd H_1}{\partial t^2}= -\nabla\times \frac{1}{\epsilon}\nabla\times \bd H_1+\nabla\times\frac{\delta}{\epsilon^2}e^{i\phi}\nabla\times\bd H_2,\\
\label{2band2}\frac{\partial^2 \bd H_2}{\partial t^2}= -\nabla\times \frac{1}{\epsilon}\nabla\times \bd H_2+\nabla\times\frac{\delta}{\epsilon^2}e^{-i\phi}\nabla\times\bd H_1.
\end{gather}
We see from Eqs. \ref{2band1}-\ref{2band2} that the two-band coupled mode equations have a sort of nonreciprocal phase $\phi$ in the coupling terms, which signals the existence of an effective gauge field. In order to identify the effective gauge field, recall that the Hamiltonian of a charged particle under a gauge field $\bd A$ is given by $\hat H=(-i\nabla-q\bd A)^2/2m=(-\nabla^2+2iq\bd A\cdot\nabla+iq\nabla\cdot\bd A+q^2\bd A^2)/2m$, where $iq\nabla\cdot\bd A+q^2\bd A^2$ is related to gauge freedom and shift of particle energy. The term $2iq\bd A\cdot\nabla$ provides the guidance to identify the gauge field which is the imaginary coefficient of the term linear in $\nabla(\cdot)$.

We manipulate Eqs. \ref{2band1}-\ref{2band2} to figure out the terms linear in $\nabla(\cdot)$ acting on $\bd H_1$. After Fourier transformation to the frequency domain and eliminating $\bd H_2$ in Eqs. \ref{2band1}-\ref{2band2} , we find the following equation of motion for $\bd H_1$
\be\label{H1f}
-\omega^2\bd H_1= -\nabla\times \frac{1}{\epsilon}\nabla\times \bd H_1-\frac{1}{\omega^2}\nabla\times\frac{\delta}{\epsilon^2}e^{i\phi}\nabla\times\nabla\times\frac{\delta}{\epsilon^2}e^{-i\phi}\nabla\times\bd H_1+\cdots,
\ee
where $\cdots$ represents higher order terms in $1/\omega^2$. For modulation frequency larger than the bandwidth of static bands, we could ignore them. Due to the same assumption, the characterizing frequency of the band of $H_1$ can be approximated by the modulation frequency $\omega$ in the rotating frame. Next, we want to extract all the terms linear in $\nabla(\cdot)$ that act on $\bd H_1$ in Eq. \ref{H1f}. We consider TE modes, i.e., $\bd H_1=H_1\bd e_z$, and the terms linear in $\nabla H_1$ are identified as follows
\bqa
&&\nabla\times\frac{\delta}{\epsilon^2}e^{i\phi}\nabla\times\nabla\times\frac{\delta}{\epsilon^2}e^{-i\phi}\nabla\times\bd H_1\\
&\sim&\frac{\delta}{\epsilon^2}e^{i\phi}\nabla^2(\nabla\frac{\delta}{\epsilon^2}e^{-i\phi}\cdot\nabla H_1)+\nabla\frac{\delta}{\epsilon^2}e^{i\phi}\cdot\nabla(\nabla\frac{\delta}{\epsilon^2}e^{-i\phi}\cdot\nabla H_1) \\
&\sim& \frac{\delta}{\epsilon^2}e^{i\phi}\nabla^2(\nabla\frac{\delta}{\epsilon^2}e^{-i\phi})\cdot\nabla H_1+\nabla\frac{\delta}{\epsilon^2}e^{i\phi}\cdot\big(\nabla(\nabla\frac{\delta}{\epsilon^2}e^{-i\phi})\cdot\nabla H_1\big),
\eqa
where in each step we have gradually ignored terms that are impossible to lead to terms linear in $\nabla H_1$. As a result,
\be
\bd A_\mathrm{eff}=\frac{1}{\omega^2}\mathrm{Im}\left\{\frac{\delta}{\epsilon^2}e^{i\phi}\nabla^2(\nabla\frac{\delta}{\epsilon^2}e^{-i\phi})+\nabla\frac{\delta}{\epsilon^2}e^{i\phi}\cdot\nabla(\nabla\frac{\delta}{\epsilon^2}e^{-i\phi})\right\}=\frac{1}{\omega^2}\mathrm{Im}\left\{\nabla\cdot\big(\frac{\delta}{\epsilon^2}e^{i\phi}\nabla(\nabla\frac{\delta}{\epsilon^2}e^{-i\phi})\big)\right\},
\ee
where the $\cdot$ in the last equality means scalar product between the first and last $\nabla$. 

Similarly, we can derive the effective gauge field for TM modes. For this purpose, we eliminate $\bd H$ fields in Eqs. \ref{CM1}-\ref{CM4} and the coupled mode equations for $\bd D$ fields are 
\begin{gather}
\label{2band1TM}\frac{\partial^2 \bd D_1}{\partial t^2}= -\nabla\times\nabla\times \frac{1}{\epsilon} \bd D_1+\nabla\times\nabla\times\frac{\delta}{\epsilon^2}e^{i\phi}\bd D_2,\\
\label{2band2TM}\frac{\partial^2 \bd D_2}{\partial t^2}= -\nabla\times\nabla\times \frac{1}{\epsilon} \bd D_2+\nabla\times\nabla\times\frac{\delta}{\epsilon^2}e^{-i\phi}\bd D_1.
\end{gather}
Eliminating $\bd D_2$ and using $\bd D_1=D_1\bd e_z$, it is straightforward to find, for TM modes, the effective gauge field is
\be
\bd A_\mathrm{eff}=\frac{2}{\omega^2}\mathrm{Im}\left\{\nabla^2(\frac{\delta}{\epsilon^2}e^{i\phi}\nabla\frac{\delta}{\epsilon^2}e^{-i\phi})\right\}.
\ee

We note for $\delta(\bd r)e^{i\phi(\bd r)}=u(\bd r) e^{i\bd q\cdot\bd r}$, where $u(\bd r)$ is a periodic function with the same periodicity as the static photonic crystal, the effective gauge fields are also periodic for both TE and TM modes, indicating there is no net effective magnetic flux in the two dimensional Floquet photonic crystals.

\section{More phase diagram of the Floquet photonic crystal}
\begin{figure}[!htb]
\begin{center}
\includegraphics[width=0.5\columnwidth]{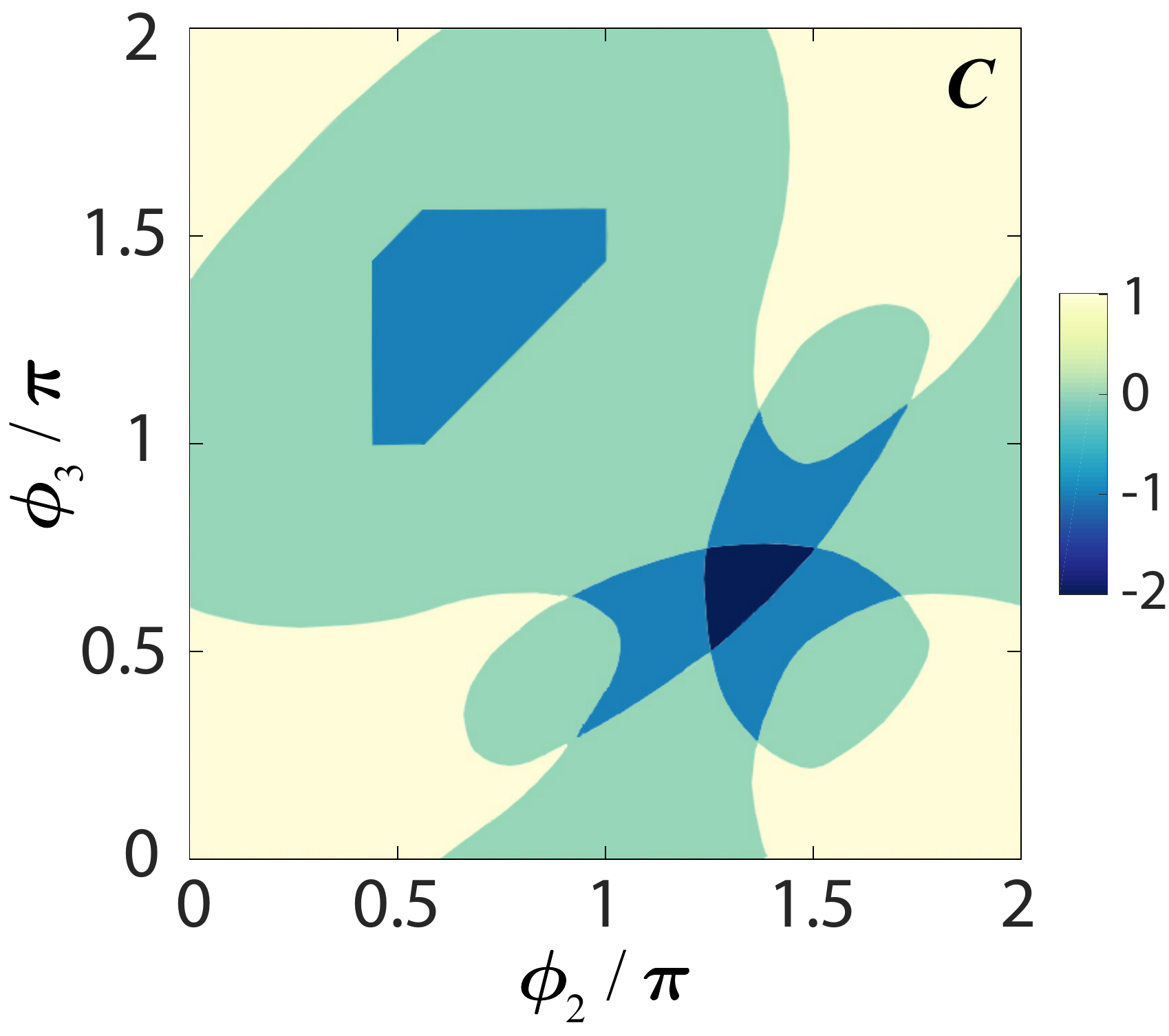}
\caption{Phase diagram of the Floquet two-band model for $\bar u(\bd r)=0.5$ in the dielectrics. } 
\label{figS1}
\end{center}
\end{figure}

The phase diagram of the Floquet photonic crystal with $\bar u(\bd r)=0.25$ in the dielectrics, as shown in Fig. 4 of the main text, has no positive Chern numbers. This changes when the modulation strength is further increased. For example, Fig. \ref{figS1} shows the phase diagram for $\bar u(\bd r)=0.5$ in the dielectrics (while other structure and modulation parameters are all the same as those in the main text). Remarkably, positive Chern number ($C=1$) shows up around $\phi_1=\phi_2=\phi_3=0$ mod $2\pi$, as it is intuitively expected to be zero when the chirality of modulation is absent.

%\noindent\textbf{Additional information}\\
%Supplementary information is available in the online version of the paper.  Reprints and permissions are available at www.nature.com/reprints.  The authors declare no competing financial interests.  Correspondence and requests for materials should be sent to OP (opainter@caltech.edu). \\
%
%\noindent\textbf{Competing financial interests}\\
%The authors declare no competing financial interests.
 
\end{document}